# COMPUTATIONAL MODELING IN SUPPORT OF NATIONAL IGNITION FACILITY OPERATIONS

M. J. Shaw, R. A. Sacks, C. A. Haynam, and W. H. Williams
Lawrence Livermore National Laboratory, Livermore, CA 94550, USA

Abstract

Numerical simulation of the National Ignition Facility (NIF) laser performance and automated control of laser setup process are crucial to the project's success. These functions will be performed by two closely coupled computer codes: the virtual beamline (VBL) and the laser operations performance model (LPOM).

## 1 INTRODUCTION

Success on many of the NIF laser's missions depends on obtaining precisely specified energy waveforms from each of the 192 beams over a wide variety of pulse lengths and temporal shapes. The LPOM automates inter-beam power balance, provides parameter checking for equipment protection and diagnostic setup, and supplies post-shot data analysis and archiving services. The LPOM relies on a NIF VBL physics model for fundamental modeling of individual beamline performance.

## 2 NIF VIRTUAL BEAMLINE

VBL is a natural consolidation and development of a number of previously distinct laser simulation efforts at Lawrence Livermore National Laboratory (LLNL). The Prop92 [1] extraction and propagation code has been the mainstay of design, verification, and component selection for the NIF laser system. The joint LLNL/Commissariat a L'Energie Atomique pumping code AMP [2] has recently been extended to model thermal deposition throughout the laser cavity. Thermal/mechanical/optical response calculations translate this heating into optical aberrations that can be applied either deterministically or stochastically to influence beam intensity statistics and focusability. The optical aberration effects of gas-path turbulence are modeled, and extension to the rest of the laser transport system has begun. Optical imperfections in the many components comprising each NIF beamline also influence beam propagation. The VBL includes these effects either as measured metrology data or as power spectral density-based simulated phase screens. The NIF adaptive optic system is critical for controlling farfield beam characteristics. Simulation played a major role in design and design-certification of this system, and this model is included in the VBL. Understanding of optical damage has progressed from early notions of threshold damage fluence through heuristic memory function treatment of pulse-shape effects [3], to today's emerging picture of stochastically distributed damage initiators coupled with thresholded exponential growth [4] of existing damage spots. This model will be included and updated as understanding develops. Conversion of optical energy from 1053-nm to 351-nm wavelength is required for NIF's scientific mission. Since the efficiency of this conversion is a major factor in NIF energetics, considerable effort has been focused on understanding and modeling this process [5]. The conversion model will be integrated into the VBL in support of the LPOM's laser-setup mission.

All facets of the VBL model are under intense development. One particular advance deals with the facility for deriving the low-power temporal shape required to generate a specified high-power output shape. For speed, the algorithm in PROP92 was based on a single-ray, energetics-only calculation. A new approach was recently implemented that uses multiple rays and an improved iterative algorithm to achieve a more accurate solution while still holding the computation time to roughly 10% of that needed for the diffractive "forward" calculation. The new method improves the solution of the input pulse solver by roughly an order of magnitude.

## 3 LPOM

The LPOM will be one of the NIF Integrated Computer Control System (ICCS) high-level software supervisors. The primary role of the LPOM is to automate the setup of the 192 individual NIF laser beams. Figure 1 illustrates LPOM's interaction with the laser system. LPOM will maintain a current model that includes the optical paths, configurations, and frequency conversion characteristics for each beam, as well as a database of diagnostic measurements, laser energy, and power at various locations along the beamline.

From these, it will estimate the amplifier gains and optical losses. These parameters will be used in VBL to calculate the laser equipment set points required to produce the requested output laser pulse. These set points include the initial energy and pulse shape produced in the Master Oscillator Room (MOR) and various attenuation settings for waveplates located in the preamplifier module (PAM) and the Preamplifier Beam Transport System (PABTS).

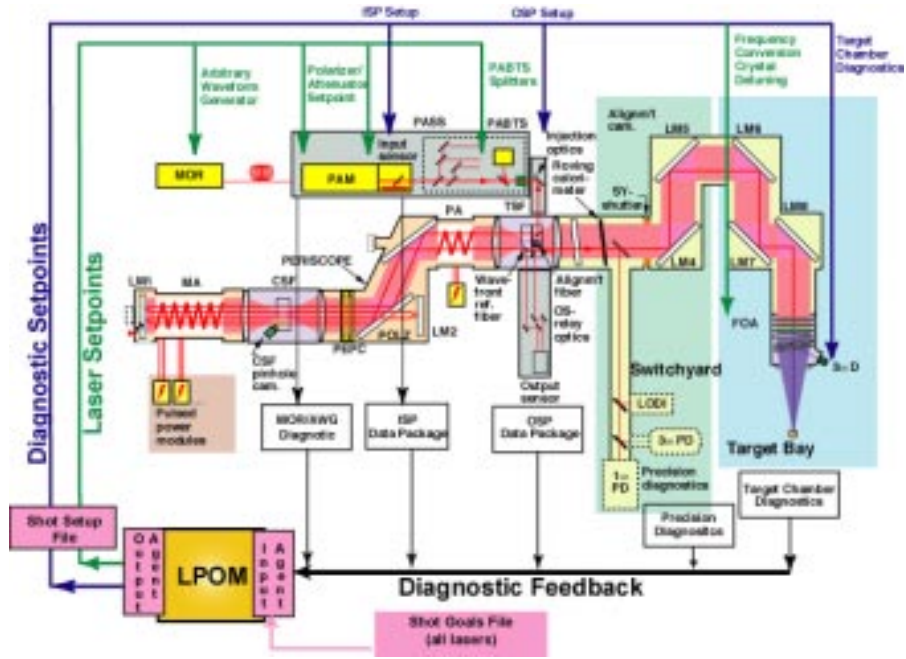

Figure 1: Laser equipment set points determined by LPOM and the diagnostic signals used for feedback. The LPOM communicates with the front-end system through a communication agent with the ICCS.

LPOM will also predict the power and energy at each of the laser diagnostic locations to aid in setting up the diagnostics system (attenuation levels, temporal range, etc.) in such a way as to optimize the clarity of the signal. Post-shot, this data is fed back to the LPOM in an automated closed-loop fashion to provide updates to the predictive model.

The LPOM Equipment Protection Module has two components. The first module, the Setup Assessment Code, checks the calculated system setup before the ICCS Laser Supervisory System implements it. In this role, LPOM will compare its calculated setpoint values against allowable ranges provided by offline VBL calculations that evaluate the danger of excessive optics damage. The second module, the Setup Verification Module, evaluates actual pulses generated and measured during a series of low-energy (~1 J) shots prior to the countdown of a full-system shot. Its role is to verify that these low-energy pulses match the expected pulse shape and energy. Both modules serve as administrative controls for equipment protection; i.e., LPOM will alert the NIF Shot Director of the status of the shot with respect to the impact upon damage-susceptible elements.

The third LPOM component is a data analysis and archiving module. This module serves two related roles. First, it provides analytical tools required to measure system performance. These include algorithms to calculate laser power balance, compare predicted laser pulses to those measured, and calculate laser system trends. Results from this module are used both for post-run evaluation and for input to future LPOM calculations. Second, it provides archiving of LPOM-generated data, user access to LPOM and its related archives, and graphical tools for trending and data reporting.

## 4 CONCEPTUAL DESIGN

A conceptual design of the LPOM system has been completed. Because of time constraints, it will be implemented on a parallel architecture, with a single processor devoted to each beamline. A master CPU will control the communication between LPOM and ICCS, coordinate laser setup and equipment protection calculations, and provide graphical user interfaces (GUIs) and data analysis tools to users.

Over the past few months, a prototype or emulation of the LPOM laser setup module has been developed.

The prototype is focused on the setup of NIF's main amplifier section and was designed to provide proof of principle of the laser setup strategy, to test parameter optimization schemes, and to develop GUIs. A PROP92 calculation with a prescribed set of amplifier gains and optical losses is used as a surrogate laser shot. A parameter optimizer uses simulated input and output laser diagnostic data (including realistic amounts of systematic and stochastic noise) to estimate a set of gains and losses that best matches the simulated data. This set of parameters, along with the Input Pulse Solver, is used to generate the input pulse shapes and energies required for subsequent shots. The emulator package is diagrammed in Figure 2.

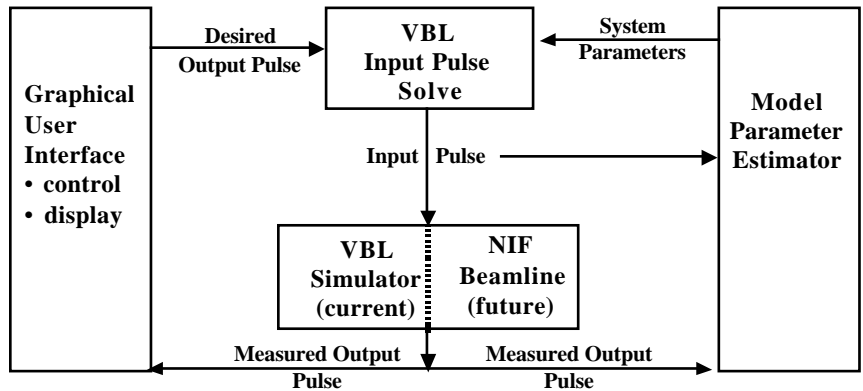

Figure 2: Conceptual design of the LPOM shot setup emulator system. The model parameter estimator consists of VMCON, using a simplified C++ laser model.

A nonlinear parameter optimization package determines the amplifier gains and optical losses of the laser as a best fit to archived diagnostic data. The parameter adjustment scheme is required to accurately track variations in the system, without introducing appreciable uncertainties in the laser output. A series of tests are under way to determine the most reliable operation strategy. In the immediate future, the emulator will be extended to include the entire laser system and then to include multiple laser beamlines.

## 5 SUMMARY

The LPOM will be an integral part of the NIF laser's supervisory control software, providing setup, equipment protection, analysis, and archiving services. It will rely on the VBL for physics modeling of the beamline performance and damage risk.

## REFERENCES

[1] R. A. Sacks et al., *1996 ICF Annual Report*, Lawrence Livermore National Laboratory, Livermore, CA, UCRL-LR-105821-96. (1996), pp. 207–213.

[2] K. S. Jancaitis et al., "A 3-dimensional ray-trace model for predicting the performance of flashlamp-pumped laser amplifiers," Proceedings of the Second Annual International Conference on Solid State Lasers for Application to Inertial Confinement Fusion, Paris, France, October 20–25, 1996.

[3] J. B. Trenholme, "Damage from Pulses with Arbitrary Temporal Shapes," Lawrence Livermore National Laboratory, Livermore, CA, memo LST-LMO-94-001 (Ver. 2) L-18179-2 (1995).

[4] C. Widmayer, M. Runkel, and M. Feit, "The Effect of LAT Beam Shape, Energy, and Pointing Fluctuations on Damage Initiation Experiments," Lawrence Livermore National Laboratory, Livermore, CA, memo NIF-0064364, April 2001.

[5] J. M. Auerbach et al., *1996 ICF Annual Report*, Lawrence Livermore National Laboratory, Livermore, CA, UCRL-LR-105821-96. (1996), pp. 199–206.